%% file: proceedings.tex
\documentclass[Physsubmission, Phys]{SciPost}

\hypersetup{
    colorlinks,
    linkcolor={red!50!black},
    citecolor={blue!50!black},
    urlcolor={blue!80!black}
}

\usepackage[bitstream-charter]{mathdesign}
\usepackage{graphicx}
\DeclareSymbolFont{usualmathcal}{OMS}{cmsy}{m}{n}
\DeclareSymbolFontAlphabet{\mathcal}{usualmathcal}

\newcommand{\f}[2]{\frac{#1}{#2}} 
\newcommand{\tf}[2]{{\textstyle \frac{#1}{#2}}} 
 
\newcommand{\de}{\partial} 
\newcommand{\la}{\langle} 
\newcommand{\ra}{\rangle}

\newcommand{\Oc}{{\cal O}} 
\renewcommand{\Re}{{\rm Re}}

\newcommand{\sgnepsilon}{\varepsilon}

\begin{document}

\begin{center}{\Large \textbf{
Lattice simulations of the QCD chiral transition at real $\mu_B$
}}\end{center}

\begin{center}
    Attila P\'asztor\textsuperscript{*1},
    Szabolcs Bors\'anyi\textsuperscript{2},
    Zolt\'an Fodor\textsuperscript{1,2,3,4,5}, 
    M. Giordano\textsuperscript{1}, 
    S. D.\ Katz\textsuperscript{1,6},
    D. N\'ogr\'adi\textsuperscript{1} and
    C. H. Wong\textsuperscript{2}
\end{center}

\begin{center}
{\bf 1} ELTE E\"otv\"os Lor\'and University, Institute for Theoretical Physics, P\'azm\'any P\'eter s\'et\'any 1/A, H-1117, Budapest, Hungary
\\
{\bf 2} Department of Physics, Wuppertal University, Gaussstr.\ 20, D-42119, Wuppertal, Germany
\\
{\bf 3} Pennsylvania State University, Department of Physics, State College, Pennsylvania 16801, USA
\\
{\bf 4} J{\"u}lich Supercomputing Centre, Forschungszentrum J{\"u}lich, D-52425 J{\"u}lich, Germany
\\
{\bf 5} Physics Department, UCSD, San Diego, CA 92093, USA 
\\
{\bf 6} MTA-ELTE Theoretical Physics Research Group, P\'azm\'any P\'eter s\'et\'any 1/A, H-1117 Budapest, Hungary
\\
* apasztor@bodri.elte.hu
\end{center}

\begin{center}
\today
\end{center}

\definecolor{palegray}{gray}{0.95}
\begin{center}
\colorbox{palegray}{
  \begin{minipage}{0.95\textwidth}
    \begin{center}
    {\it  XXXIII International (ONLINE) Workshop on High Energy Physics \\“Hard Problems of Hadron Physics:  Non-Perturbative QCD \& Related Quests”}\\
    {\it November 8-12, 2021} \\
    \doi{10.21468/SciPostPhysProc.?}\\
    \end{center}
  \end{minipage}
}
\end{center}

\section*{Abstract}
{\bf
Most lattice studies of hot and dense QCD matter rely on extrapolation from zero or imaginary chemical potentials. 
The ill-posedness of numerical analytic continuation puts severe limitations on the reliability of such methods. We 
studied the QCD chiral transition at finite real baryon density with the more direct sign reweighting approach. 
We simulate up to a baryochemical potential-temperature ratio of $\mu_B/T=2.7$, covering the RHIC Beam Energy Scan range, and penetrating the region
where methods based on analytic continuation are unpredictive.
This opens up a new window to study QCD matter at finite $\mu_B$ from first principles. This conference contribution is based on Ref.~\cite{Borsanyi:2021hbk}. 
}

\vspace{10pt}
\noindent\rule{\textwidth}{1pt}
\tableofcontents\thispagestyle{fancy}
\noindent\rule{\textwidth}{1pt}
\vspace{10pt}

\section{Introduction}
\subsection{QCD at finite $\mu_B$ and the need for more direct methods}
One of the major unsolved problems in high energy physics is the calculation of the phase diagram 
of strongly interacting matter in the temperature ($T$) - baryochemical potential ($\mu_B$) plane.
Many aspects of QCD thermodynamics at $\mu_B=0$ have been clarified by first principle lattice QCD 
calculations, such as the crossover nature of the transition and the value of the transition 
temperature~\cite{Aoki:2006we,Borsanyi:2010bp,Bazavov:2011nk}. 

It is conjectured that at higher baryochemical potential the QCD crossover
gets stronger and above a certain point turns into a first order phase transition. 
The endpoint of the line of first order transitions is called the critical endpoint.
Establishing the existence and the location of this conjectured critical endpoint is one
of the main goals of the phenomenology of heavy ion collisions and of QCD thermodynamics.

First principle lattice calculations at finite $\mu_B$ are, however, hampered by the notorious complex-action problem: the path integral
weights become complex numbers, and importance sampling breaks down. A number of methods have been 
introduced over the years to side-step this problem. In particular, most state-of-the-art calculations
involve analytic continuation using either i) data on Taylor coefficients of different observables at $\mu_B=0$ or 
ii) data at purely imaginary chemical potentials $\mu_B^2 \leq 0$, where the sign problem is absent. An example of an
important result coming from these approaches is the calculation of the curvature of the crossover line $T_c(\mu_B)$ near 
zero chemical potential~\cite{HotQCD:2018pds,Bonati:2018nut,Borsanyi:2020fev}. Another important result is the
calculation of the Taylor coefficients of the pressure in a series expansion in
the chemical potential up to fourth order~\cite{Bellwied:2015lba, Bazavov:2017dus}, which 
have been calculated on the lattice up to high enough temperatures to match results from resummed perturbation theory~\cite{Mogliacci:2013mca,Haque:2014rua}.

The extension of these results to higher orders in the Taylor expansion and to higher chemical potentials, however, faces immense challenges:
For the Taylor method, the signal-to-noise ratio increases significantly with increasing order of the Taylor expansions.
Similarly, in the determination of the same high-order coefficients with the imaginary chemical potential method, one runs 
into the ill-posedness of high-order numerical differentiation. Even if the high-order coefficients were available, 
extrapolation by a Taylor polynomial ansatz is limited by the radius of convergence of such an expansion.
While there were attempts to locate the leading singularity of the pressure with several different 
methods~\cite{Giordano:2019slo,Giordano:2019gev,Mondal:2021jxk,Dimopoulos:2021vrk}, these calculations have so
far not reached the continuum limit. Even if one knows the leading singularity determining the 
radius of convergence, it is not obvious how to go beyond it. Several resummation schemes have been experimented with, 
including Padé resummation in Refs.~\cite{Lombardo:2005ks,Pasztor:2020dur,Dimopoulos:2021vrk}, a joint expansion in temperature and 
chemical potential along lines of constant physics in Ref.~\cite{Borsanyi:2021sxv}, and a truncated reweighting scheme in Ref.~\cite{Mondal:2021jxk}.
While these methods are interesting, at the moment they provide no clear way of going beyond the crossover region of the conjectured phase diagram.
Moreover, these type of reweighting schemes have so far been used mostly to calculate observables that are not very sensitive to criticality - such as 
the pressure and the transition line $T_c(\mu_B)$. Extrapolations of observables that are sensitive to criticality, such as the width of the 
transition, are even less under control~\cite{Borsanyi:2020fev}.

To shed light on the ultimate fate of the QCD crossover at finite $\mu_B$, it is therefore of great importance to come up with more direct methods, that 
can provide results directly at a finite chemical potential, and are free of additional systematic effects, such as the aforementioned analytic continuation
problem of the Taylor and imaginary chemical potential methods, or the convergence issues of 
complex Langevin~\cite{Parisi:1983mgm,Aarts:2009uq, Sexty:2013ica}.

\subsection{Reweighting and the overlap problem}
\label{sec:intro}
Given a theory with fields $U$, reweighting is a general strategy to calculate expectation values in a target theory - with 
path integral weights $w_t$ and partition function $Z_t = \int \mathcal{D} U w_t(U)$ - 
by performing simulations in a different (simulated) theory - with path integral weights $w_s$ and partition
function $Z_s=\int \mathcal{D} U w_s(U)$. The ratio of the partition functions and expectation value in 
the target theory are then given by
\begin{equation}
  \label{eq:reweight}
    \f{Z_t}{Z_s} = \left\langle \f{w_t}{w_s}\right\rangle_{s} \quad \textrm{and} \quad
  \left\langle \Oc \right\rangle_t = \frac{\left\langle \frac{w_t}{w_s}
      \Oc \right\rangle_s }{\left\langle
      \frac{w_t}{w_s}\right\rangle_s}\,
\end{equation}
respectively, where $\left\langle \dots \right\rangle_{t,s}$ denotes taking expectation value with respect to the 
weights $w_t$ and $w_s$, respectively. In the present conference contribution, we will consider examples where the
target theory is QCD at finite baryochemical potential discretized on the lattice. 
In this case the partition function of the target theory is:
\begin{equation}
    Z_\mu = \int \mathcal{D}U \det M(U,\mu,m) e^{-S_g(U)} = \int \mathcal{D}U \operatorname{Re} \det M(U,\mu,m) e^{-S_g(U)}\rm{,}
\end{equation}
where $S_g$ is the gauge action, $\det M$ denotes the fermionic
determinant, including all quark types with their respective masses collectively denoted by $m$,
their respective chemical potentials collectively denoted by $\mu$, as well as rooting in the case 
of staggered fermions, and the integral is over all link variables $U$. Replacing the determinant
with its real part is not permitted for arbitrary expectation values,
but it is allowed for i) observables satisfying either $\Oc(U^*) = \Oc(U)$ or ii)
observables obtained as derivatives of $Z$ with respect to real
parameters, such as the chemical potential, the quark mass or the gauge coupling. 

Since the target theory is lattice QCD at finite chemical potential,
the weights $w_t$ have wildly fluctuating phases: this is
the infamous sign problem of lattice QCD at finite baryon density. 
In addition to this problem,
generic reweighting methods also suffer from an overlap problem: the
probability distribution of the reweighting factor $w_t/w_s$ has
generally a long tail, which cannot be sampled efficiently 
in standard Monte Carlo simulations.

Many attempts at reweighting to finite baryochemical potential, such as Refs.\cite{Hasenfratz:1991ax,Fodor:2001au,Fodor:2004nz,Giordano:2019gev} 
use reweighting from zero chemical potential, when the weights are proportional to the ratio of determinants $\det M(\mu) / \det M(0)$. 
However, these studies have so far been restricted to coarse lattices, with temporal extent $N_\tau=4$, and mostly an unimproved staggered action, 
with the exception of Ref.~\cite{Giordano:2019gev}, that uses the 2stout improved staggered action~\cite{Borsanyi:2010bp}, albeit still at
$N_\tau=4$. It was actually demonstrated in Ref.~\cite{Giordano:2020uvk}, that the main bottleneck in extending such studies to finer lattices is 
the overlap problem in the weights $w_t/w_s$, which becomes severe already at moderate chemical potentials, where the sign problem is still
numerically manageable.

This overlap problem in the weights $w_t/w_s$ is not present if they take values in a compact space. The most well-known 
of these approaches is phase reweighting~\cite{Fodor:2007vv,Endrodi:2018zda}, where the simulated theory - the so called phase quenched theory - has path integral weights: 
\begin{equation}
    w_s=w_{PQ} = |\operatorname{det} M_{ud}(\mu)^{\frac{1}{2}}| \operatorname{det} M_s(0)^{\frac{1}{4}} e^{-S_g}\rm{.} 
\end{equation}
In this case the reweighting factors are pure phases:
\begin{equation}
    \left( \f{w_t}{w_s} \right)_{PQ} = e^{i \theta}\rm{,}
\end{equation}
where $\theta = \operatorname{Arg} \det M$. 
We will contrast this approach with sign reweighting, where the simulated - sign quenched - ensemble has weights:
\begin{equation}
    w_s=w_{SQ} = |\operatorname{Re} \operatorname{det} M_{ud}(\mu)^{\frac{1}{2}}| \operatorname{det} M_s(0)^{\frac{1}{4}} e^{-S_g}\rm{.} 
\end{equation}
In this case the reweighting factor are signs:
\begin{equation}
    \left( \f{w_t}{w_s} \right)_{SQ} = \epsilon \equiv \operatorname{sign} \cos \theta = \pm1\rm{,}
\end{equation}
provided that the target theory is the one with $w_t = \operatorname{Re} \det M e^{-S_g}$, i.e., provided one restricts one's attention to observables satisfying i) or ii). 

\section{The severity of the sign problem}

A measure of the strength of the sign problem in the phase reweighting scheme is given by the expectation 
value of the phases $\f{Z_{\mu}}{Z_{PQ}}=\left \langle \cos \theta \right \rangle_{PQ}$. Similarly, in the sign reweighting scheme
the severity of the sign problem is measured by $\f{Z_\mu}{Z_{SQ}}=\left \langle \epsilon \right\rangle_{SQ}$. 
The earliest mention of the sign reweighting approach we are aware of is Ref.~\cite{deForcrand:2002pa}, where it was noted that 
out of the reweighting schemes where the weights $w_t/w_s$ are a function of the phase of the quark determinant only, sign reweighting is 
the optimal one, with the weakest sign problem, in the sense that the ratio $Z_t/Z_s$ is maximal. In this section we study how much 
one gains by this optimality property of the sign quenched ensemble, when compared to the phase quenched ensemble. For this purpose we introduce 
a simplified model - to be later compared with direct simulation data - where the distribution of the phases $\theta$ in the phase quenched ensemble is given by a wrapped Gaussian distribution:
\begin{equation}
    P_{\rm PQ}(\theta) \underset{\substack{\text{Gaussian}\\\text{approx.}}}{=} \f{1}{\sqrt{2\pi}\sigma}\sum_{n=-\infty}^\infty  e^{-\f{1}{2\sigma^2}(\theta + 2\pi n )^2} \rm{.} 
\end{equation}
Once one has a model for this probability distribution, the strength of the sign problem can be estimated in both the phase and sign quenched ensembles. The estimates and 
their small chemical potential (i.e., small $\sigma$) asymptotics are given by:
\begin{equation}
    \begin{aligned}
        \la \cos \theta\ra^{\rm PQ}_{T,\mu} &= e^{-\f{\sigma^2(\mu)}{2}} \underset{\mu_B\to  0}{\sim} 1 - \f{\sigma^2(\mu)}{2}\,, \\
        \la \sgnepsilon\ra^{\rm SQ}_{T,\mu} &= \f{\la \cos \theta\ra^{\rm PQ}_{T,\mu}}{\la \left| \cos \theta \right| \ra^{\rm PQ}_{T,\mu}}
    \underset{\mu_B\to  0}{\sim}
  1 -\left(\tf{4}{\pi} \right)^{\f{5}{2}}
  \left(\tf{\sigma^2(\mu)}{2}\right)^{\f{3}{2}} e^{-\f{\pi^2}{8
      \sigma^2(\mu)}}\,. \\
    \end{aligned}
\end{equation}
Note the two very different asymptotics at small chemical potential: the phase reweighting approach leads to a regular Taylor series, while 
in the sign reweighting approach the asymptotics approach 1 faster than any polynomial.  

The large-$\mu$ or large volume
asymptotics are on the other hand very similar: in the large-$\sigma$
limit a wrapped Gaussian tends to the uniform distribution, and so at
large chemical potential or volume one arrives at
\begin{equation}
  \label{eq:asyratio}
  \f{\la \sgnepsilon\ra^{\rm SQ}_{T,\mu}}{\la \cos\theta\ra^{\rm PQ}_{T,\mu}}
  \underset{\mu_B\,\text{or}\, V\to \infty}{\sim}
  \left(
    \int_{-\pi}^{\pi}d\theta\, |\!\cos \theta| \right)^{-1}=\frac{\pi}{2} \,, 
\end{equation}
which asymptotically translates to a factor of
$(\f{\pi}{2})^2 \approx 2.5$ less statistics needed for a sign
quenched as compared to a phase quenched simulation.  

To have a numerical estimate of the strength of the
sign problem as a function of $\mu$, rather than $\sigma$ we further approximate the variance of the weights by the leading order Taylor expansion~\cite{Allton:2002zi}:
\begin{equation}
\sigma(\mu)^2 \approx \left\langle \theta^2 \right\rangle_{\rm LO}
    =-\frac{4}{9} \chi^{ud}_{11} \left( LT \right)^3 \left( \f{\mu_B}{T} \right)^2 \rm{,}
\end{equation}
where 
\begin{equation}
    \chi^{ud}_{11} = \f{1}{T^2}\frac{\partial^2 p }{\partial
  \mu_u \partial \mu_d
  }|_{\mu_u=\mu_d=0}
\end{equation}
is the disconnected part of the light
quark susceptibility, which is easily obtained by performing 
simulations at zero chemical potential. 

\begin{figure*}[t]
  \centering
  \includegraphics[width=0.45\textwidth]{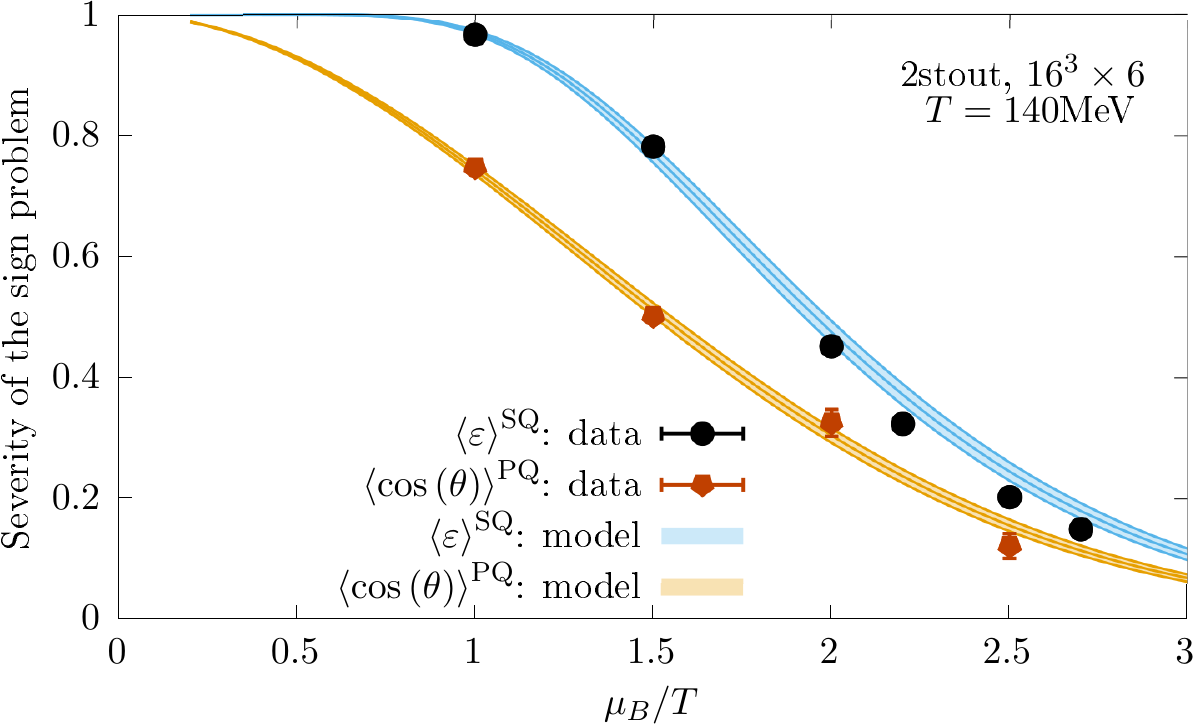}
  \includegraphics[width=0.45\textwidth]{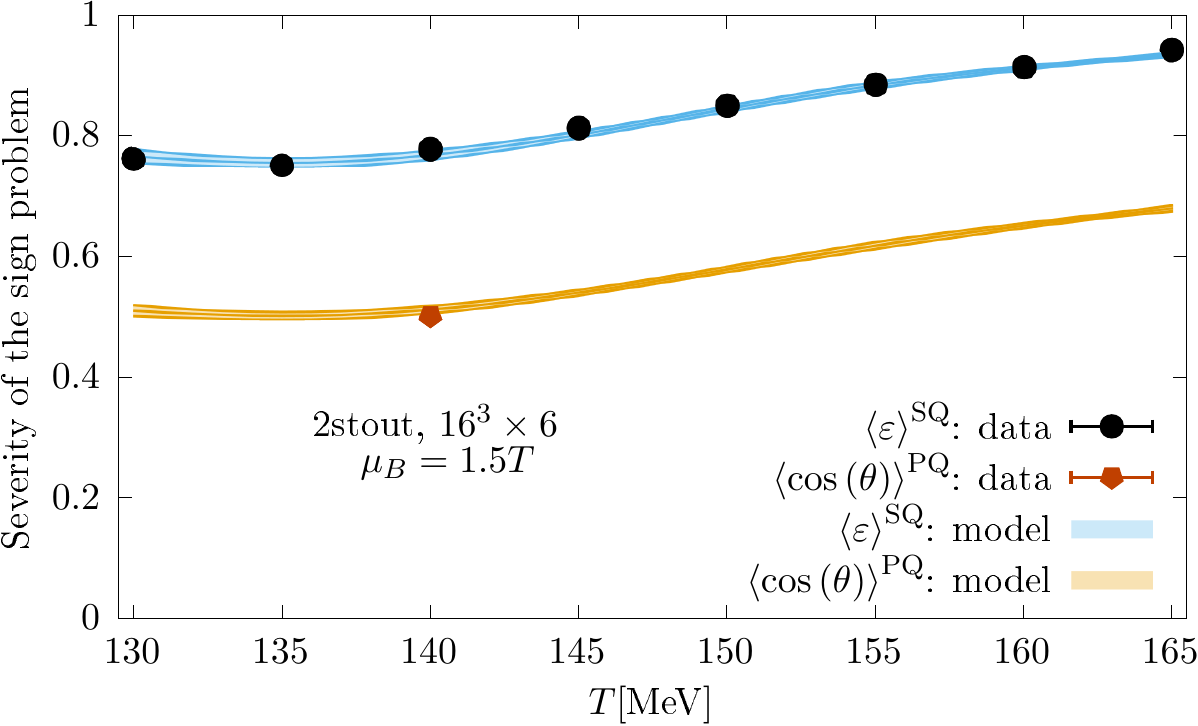}
  \caption{The strength of the sign problem on 2stout improved $16^3 \times 6$ staggered lattices 
    as a function of $\mu_B/T$
    at $T=140$ MeV (left) and as a function of $T$ at $\mu_B/T=1.5$. A
    value close to $1$ shows a mild %weak
    sign problem, while a small value
    indicates a severe sign problem.  Data for sign reweighting
    (black) and phase reweighting (orange) are from
    simulations.  Predictions of the Gaussian model (see text) are also shown.}
  \label{fig:sign}
\end{figure*}

The simple approximations made above are actually quite close to the actual simulation data,
as can be seen in Fig.~\ref{fig:sign}: our simple model predicts the strength of the sign problem
both as a function of $\mu_B$ at a fixed temperature (left) and as a function of temperature at a 
fixed $\mu_B/T$ (right). 
While deviations are
visible at larger $\mu$, even at the upper end of our $\hat{\mu}_B \equiv \frac{\mu_B}{T}$
range the deviation is at most $25\%$, and Eq.~\eqref{eq:asyratio}
approximates well the relative severity of the sign problem in the two
ensembles at $\mu_B/T > 1.5$.
This is of great practical 
importance, as it makes the planning of future simulation projects
with either the sign or phase reweighting approaches relatively straightforward: simulation costs can be
easily estimated beforehand.

\section{Lattice setup and numerical results}
For the simulations we used a tree level Symanzik improved gauge action with
the staggered Dirac operator being a function of fat links, obtained by two steps 
of stout smearing~\cite{Morningstar:2003gk} with parameter $\rho=0.15$. We only 
introduce a chemical potential for the up and down quarks, that have the same
chemical potential $\mu=\mu_l=\mu_u=\mu_d=\mu_B/3$, while for the strange quark we 
have $\mu_s=0$. We used a lattice size of $16^3\times 6$,
and performed a scan in chemical potential
at fixed $T=140\,{\rm MeV}$, and a scan in temperature at fixed
$\mu_B/T= 1.5$.  In both cases, simulations were performed by modifying the RHMC
algorithm at $\mu_B=0$ by including an extra accept/reject step that
takes into account the factor $\f{|\Re \det M_{ud}(\mu)^{\frac{1}{2}}|}{\det M_{ud}(0)}$.  The
determinant was calculated with the reduced matrix
formalism~\cite{Hasenfratz:1991ax} and dense linear algebra, with no
stochastic estimators involved. 

The main observables we studied were the light quark condensate and density.
The light-quark chiral condensate was obtained via the formula
\begin{equation}
  \label{eq:chiralcondensate}
  \begin{aligned}
    \la\bar{\psi}\psi\ra_{T,\mu} &= \f{1}{Z(T,\mu)}\f{\de
      Z(T,\mu)}{\de
      m_{ud}} 
    =\f{T}{V}\f{1}{\la\sgnepsilon\ra^{\rm SQ}_{T,\mu}}\left\la
      \sgnepsilon\f{\de}{\de m_{\rm ud}} \ln\left |\Re \det M_{ud}^{\f{1}{2}}\right|
    \right\ra^{\rm SQ}_{T,\mu} \,,
  \end{aligned}
\end{equation}
using a numerical differentiation of the determinant $\det M=\det M(U,m_{ud},m_s,\mu)$ calculated with
the reduced matrix formalism of Ref.~\cite{Hasenfratz:1991ax}.
The step size in the derivative was chosen small enough to make the systematic error 
from the finite difference negligible compared to the statistical error.  The
additive and multiplicative divergences in the condensate were renormalized with the prescription
\begin{equation}
  \label{eq:chiralcondensate3}
  \la \bar{\psi}\psi\ra_R(T,\mu) =   -\f{m_{ud}}{f_\pi^4}\left[
    \la \bar{\psi}\psi\ra_{T,\mu} -\la \bar{\psi}\psi\ra_{0,0}
  \right]\,.
\end{equation}
We also calculated the light quark density 
\begin{equation}
\begin{aligned}
  \chi^l_1 &\equiv \f{\partial \left(p/T^4\right)}{\partial \left(
      \mu/T \right) } = \frac{1}{VT^3} \f{1}{Z(T,\mu)} \f{\partial
    Z(T,\mu)}{\partial \hat{\mu}}  %\,.
  = \f{1}{VT^3 \la\sgnepsilon\ra^{\rm SQ}_{T,\mu}} \left\la
    \sgnepsilon\f{\de}{\de \hat{\mu}} \ln\left |\Re \det M_{ud}^{\f{1}{2}}\right|
  \right\ra^{\rm SQ}_{T,\mu} \,.
\end{aligned}
\end{equation}
In this case the derivative on a fixed configuration can be obtained analytically using the reduced matrix
formalism. The light quark density does not have to be renormalized.

\begin{figure*}[t]
  \centering
  \includegraphics[width=0.45\textwidth]{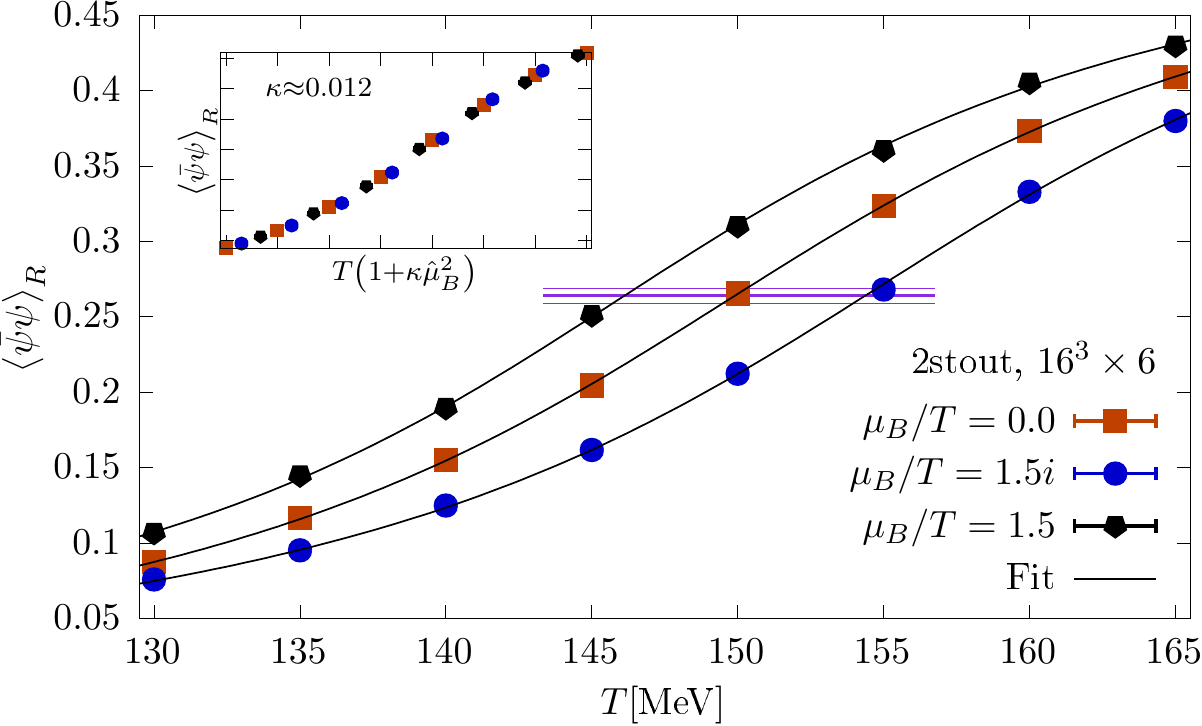}
  \includegraphics[width=0.45\textwidth]{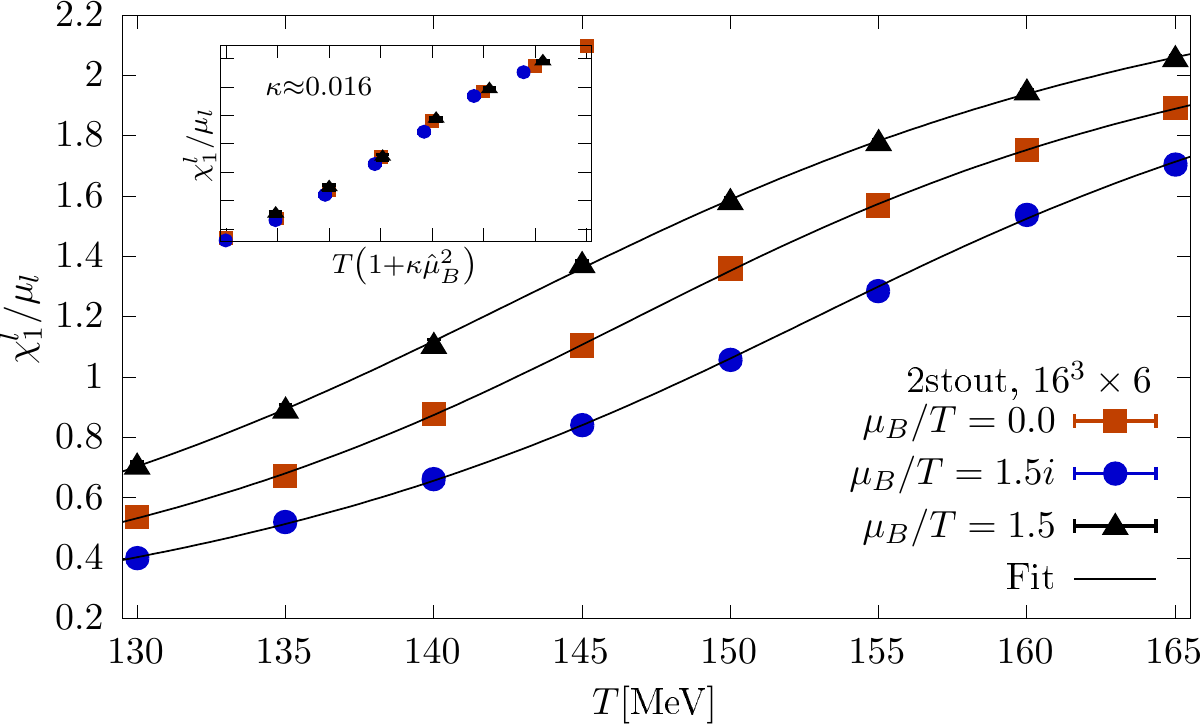}
  \caption{The renormalized chiral condensate (left) and the light
    quark number-to-light quark chemical potential ratio (right) as a
    function of $T$ at fixed $\mu_B/T=1.5,0$ and $1.5 i$ on 2stout mproved 
    lattices at $N_\tau=6$. The insets show a rescaling of the temperature 
    axis by $T \to T\left( 1 + \kappa \left( \frac{\mu_B}{T} \right)^2 \right)$,
    which approximately collapses the curves onto each other if $\kappa \approx 0.012$
    and $0.016$ are chosen for the chiral condensate and the quark number-to-chemical 
    potential ratio, respectively.
    }
  \label{fig:Tscan}
\end{figure*}

\begin{figure*}[t]
  \centering
  \includegraphics[width=0.45\textwidth]{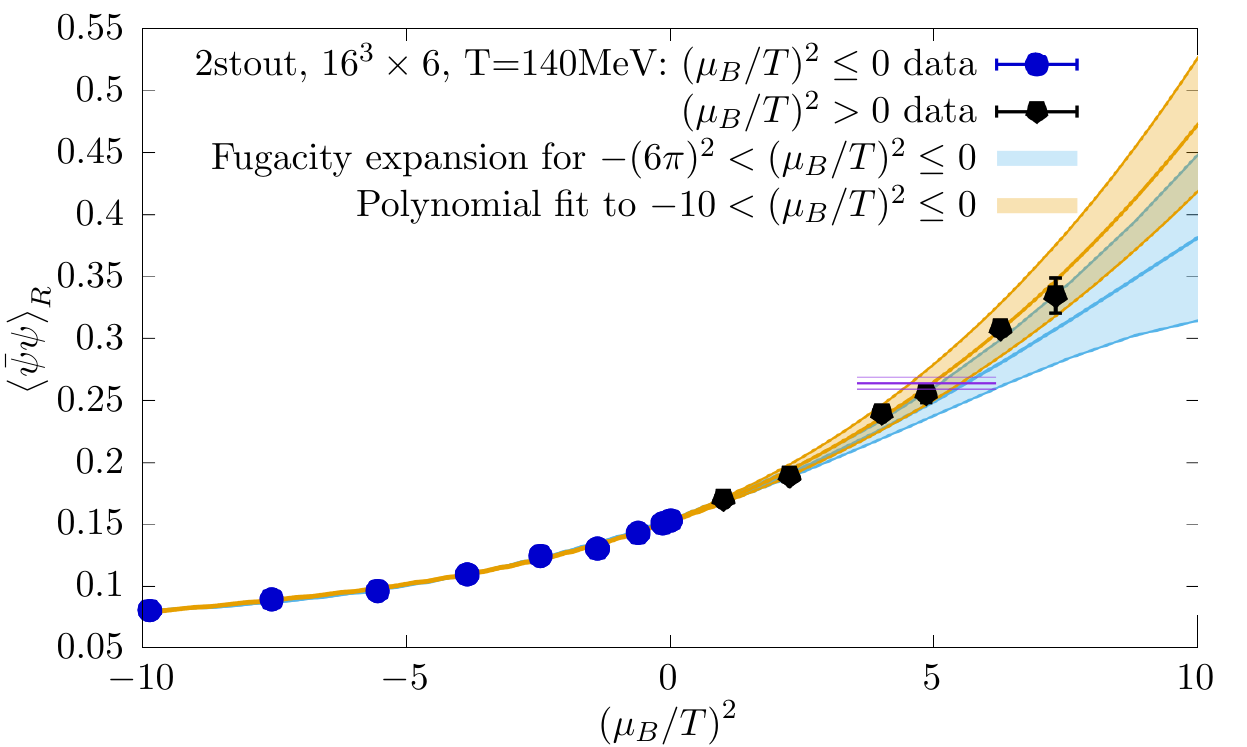}
  \includegraphics[width=0.45\textwidth]{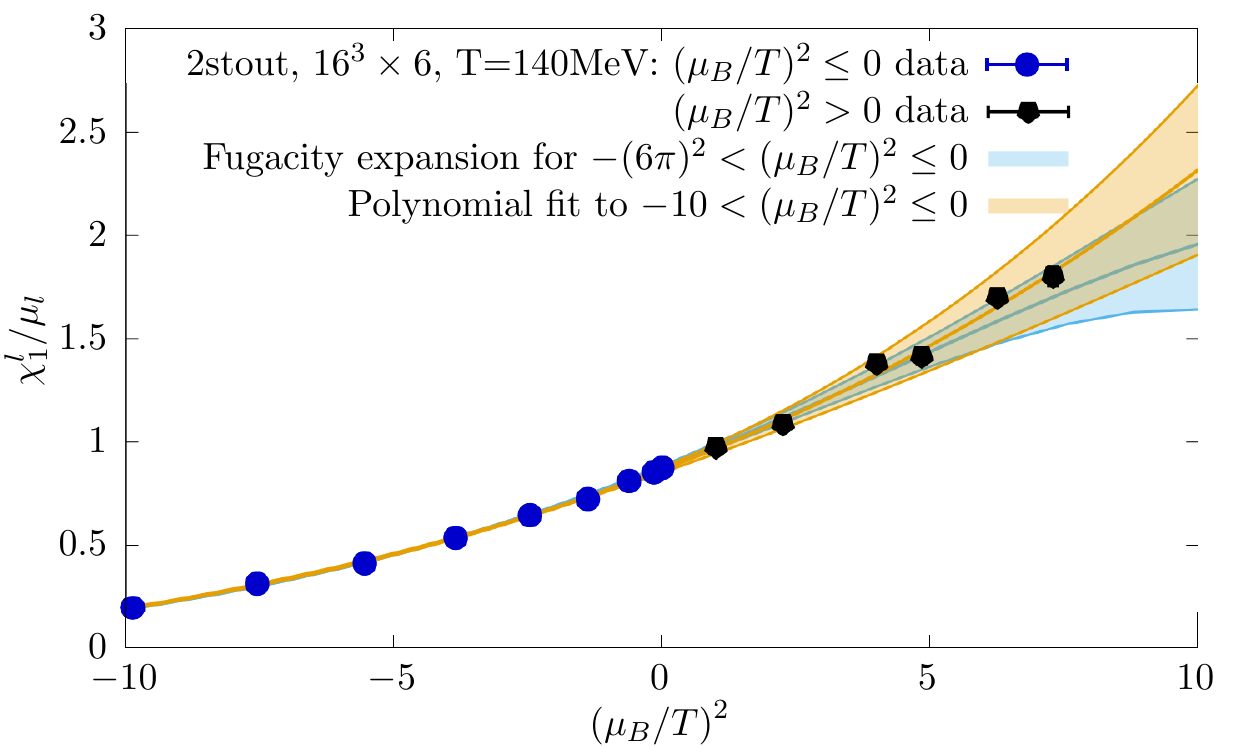}
  \caption{The renormalized chiral condensate (left) and the light
    quark number-to-light quark chemical potential ratio (right) as a
    function of $\left( \mu_B/T \right)^2$ at temperature $T=140$
    MeV with the 2stout improved staggered action at $N_\tau=6$. Data from simulations at real $\mu_B$ (black) are compared
with analytic continuation from simulations at imaginary $\mu_B$ (blue).
    In the
    left panel the value of the condensate at the crossover
    temperature at $\mu_B=0$ is also shown by the horizontal line.  
    The simulation data cross
    this line at $\mu_B/T \approx 2.2$.
    }
  \label{fig:muscan}
\end{figure*}

Our results for a temperature scan between $130$ MeV and $165$ MeV at
real chemical potential $\mu_B/T=1.5$, zero chemical potential,
and imaginary chemical potential $\mu_B/T=1.5i$ are shown in
Fig.~\ref{fig:Tscan}. We also show that a
rescaling of the temperature axis of the form $T \to T\left(1+\kappa \left( \frac{\mu_B}{T} \right)^2\right)$,
where $\kappa \approx 0.012$ for the chiral condensate and $\kappa \approx 0.016$ for $\chi^l_1/\mu_l$
collapses the curves into each other. Such a simple rescaling indicates that up to $\mu_B/T=1.5$ the chiral crossover
does not get narrower, which is what one would expect in the vicinity of a critical endpoint. 

Our results for the chemical potential scan at a fixed temperature of
$T=140$ MeV are shown in Fig.~\ref{fig:muscan}.  We have performed
simulations at 
$\mu_B/T=1, 1.5, 2, 2.2, 2.5, 2.7$.
The sign-quenched results are compared with the results of 
analytic continuation from imaginary chemical potentials.
To demonstrate the magnitude of the systematic errors of such an 
extrapolation we considered two fits.  (\textit{i}) As the simplest
ansatz, we fitted the data with a cubic polynomial in
$\hat{\mu}_B^2 = \left( \f{\mu_B}{T} \right)^2$
in the range
$\hat{\mu}_B^2 \in [-10,0]$.
(\textit{ii}) As an alternative, we
also and ans\"atze for both
$\left\langle \bar{\psi}\psi\right\rangle_R$ and
$\chi^l_1/\hat{\mu}_l$ based on the fugacity expansion
$p/T^4 = \sum_n A_n \cosh(n \mu_l/T)$, fitting the data in the
entire imaginary-potential range
$\hat{\mu}_B^2 \in \left[ -(6 \pi)^2,0\right]$
using respectively 7 and
6 fitting parameters.  Fit results are also shown in
Fig.~\ref{fig:muscan}; only statistical errors are displayed.  While
sign reweighting and analytic continuation 
give compatible results, in the upper half of the
$\mu_B$ range the errors from sign reweighting are an order of
magnitude smaller.  In fact, sign reweighting can penetrate the region
$\hat{\mu}_B>2$ where the extrapolation of many quantities is not yet
possible with standard methods~\cite{Bazavov:2017dus,Borsanyi:2020fev}.

\section{Conclusions}
Due to the increasing computing power of modern hardware, direct approaches to finite density QCD are becoming increasingly
feasible, and are opening up a new window to study the bulk thermodynamics of strongly interacting matter from first principles.
In this conference contribution and the paper Ref.~\cite{Borsanyi:2021hbk} which it is based on, we studied the method of sign reweighting in 
detail for the first time. 
While the method is ultimately bottlenecked by the sign problem, in the region of applicability it offers excellent reliability
compared to the dominant methods of Taylor expansion and imaginary chemical potentials - which always provide results having a shadow of a doubt
hanging over them due to the analytic continuation problem. We have demonstrated that the strength of the sign problem can be easily estimated
with $\mu=0$ simulations, making the method practical and the planning of simulation projects straightforward. We have also demonstrated that
the method extends well into the regime where the established methods start to lose predictive power, and covers the range of the RHIC Beam Energy Scan (BES)~\cite{STAR:2017sal,STAR:2020tga}. 

The lattice action used in this study is often 
the first point
of a continuum extrapolation in QCD thermodynamics.  Furthermore,
while the sign problem is exponential in the physical volume, it is
not so in the lattice spacing. Continuum-extrapolated finite $\mu_B$
results in the range of the RHIC BES and is already
within reach for the phenomenologically relevant aspect ratio of
$LT \approx 3$. 

On a more methodological point, the phase and sign reweighting approaches only guarantee the absence of heavy tailed distributions
when calculating the ratio of the partition functions (or the pressure difference) of the target and 
simulated theories. 
Furthermore, the optimum property of the sign quenched ensemble is only a statement about the 
denominator of Eq.~\eqref{eq:reweight} (right). The optimal ensemble when both the numerator and the 
denominator are taken into account is most likely, however, observable dependent. 
For these two reasons, the study of the probability distributions of observables other than the pressure 
is an important direction for future work.

\section*{Acknowledgements}
The project was supported by the BMBF Grant No. 05P18PXFCA.
This work was also supported by the Hungarian National Research,
Development and Innovation Office, NKFIH grant KKP126769.
A.P. is supported by the J. Bolyai Research
Scholarship of the Hungarian Academy of Sciences and by the \'UNKP-21-5 New
National Excellence Program of the Ministry for Innovation and Technology
from the source of the National Research, Development and Innovation Fund.
The authors gratefully acknowledge the Gauss Centre for Supercomputing e.V.
(www.gauss-centre.eu) for funding this project by providing computing time on the
GCS Supercomputers JUWELS/Booster and JURECA/Booster at FZ-Juelich.

%\bibliographystyle{SciPost_bibstyle} % Include this style file here only if you are not using our template
%\bibliography{thermo.bib}
\input{proceedings.bbl}

\nolinenumbers

\end{document}

%% file: proceedings.bbl
\providecommand{\noopsort}[1]{}\providecommand{\singleletter}[1]{#1}%

%% file: proceedings.bbl
\begin{thebibliography}{10}
\providecommand{\url}[1]{\texttt{#1}}
\providecommand{\urlprefix}{URL }
\expandafter\ifx\csname urlstyle\endcsname\relax
  \providecommand{\doi}[1]{doi:\discretionary{}{}{}#1}\else
  \providecommand{\doi}{doi:\discretionary{}{}{}\begingroup
  \urlstyle{rm}\Url}\fi
\providecommand{\eprint}[2][]{\url{#2}}

\bibitem{Borsanyi:2021hbk}
S.~Bors\'anyi, Z.~Fodor, M.~Giordano, S.~D. Katz, D.~Nogr\'adi, A.~P\'asztor
  and C.~H. Wong,
\newblock \emph{{Lattice simulations of the QCD chiral transition at real
  baryon density}}  (2021),
\newblock \eprint{2108.09213}.

\bibitem{Aoki:2006we}
Y.~Aoki, G.~Endr{\H o}di, Z.~Fodor, S.~D. Katz and K.~K. Szab{\'o},
\newblock \emph{{The Order of the quantum chromodynamics transition predicted
  by the standard model of particle physics}},
\newblock Nature \textbf{443}, 675 (2006),
\newblock \doi{10.1038/nature05120},
\newblock \eprint{hep-lat/0611014}.

\bibitem{Borsanyi:2010bp}
{\relax Sz}.~Bors{\' a}nyi, Z.~Fodor, C.~Hoelbling, S.~D. Katz, S.~Krieg,
  C.~Ratti and K.~K. Szab{\'o},
\newblock \emph{{Is there still any $T_c$ mystery in lattice QCD? Results with
  physical masses in the continuum limit III}},
\newblock JHEP \textbf{09}, 073 (2010),
\newblock \doi{10.1007/JHEP09(2010)073},
\newblock \eprint{1005.3508}.

\bibitem{Bazavov:2011nk}
A.~Bazavov \emph{et~al.},
\newblock \emph{{The chiral and deconfinement aspects of the QCD transition}},
\newblock Phys. Rev. D \textbf{85}, 054503 (2012),
\newblock \doi{10.1103/PhysRevD.85.054503},
\newblock \eprint{1111.1710}.

\bibitem{HotQCD:2018pds}
A.~Bazavov \emph{et~al.},
\newblock \emph{{Chiral crossover in QCD at zero and non-zero chemical
  potentials}},
\newblock Phys. Lett. B \textbf{795}, 15 (2019),
\newblock \doi{10.1016/j.physletb.2019.05.013},
\newblock \eprint{1812.08235}.

\bibitem{Bonati:2018nut}
C.~Bonati, M.~D'Elia, F.~Negro, F.~Sanfilippo and K.~Zambello,
\newblock \emph{{Curvature of the pseudocritical line in QCD: Taylor expansion
  matches analytic continuation}},
\newblock Phys. Rev. D \textbf{98}(5), 054510 (2018),
\newblock \doi{10.1103/PhysRevD.98.054510},
\newblock \eprint{1805.02960}.

\bibitem{Borsanyi:2020fev}
{\relax Sz}.~Bors{\' a}nyi, Z.~Fodor, J.~N. G{\"u}nther, R.~Kara, S.~D. Katz,
  P.~Parotto, A.~P{\'a}sztor, C.~Ratti and K.~K. Szab{\'o},
\newblock \emph{{QCD Crossover at Finite Chemical Potential from Lattice
  Simulations}},
\newblock Phys. Rev. Lett. \textbf{125}(5), 052001 (2020),
\newblock \doi{10.1103/PhysRevLett.125.052001},
\newblock \eprint{2002.02821}.

\bibitem{Bellwied:2015lba}
R.~Bellwied, S.~Bors{\'a}nyi, Z.~Fodor, S.~D. Katz, A.~P{\'a}sztor, C.~Ratti
  and K.~K. Szab{\'o},
\newblock \emph{{Fluctuations and correlations in high temperature QCD}},
\newblock Phys. Rev. D \textbf{92}(11), 114505 (2015),
\newblock \doi{10.1103/PhysRevD.92.114505},
\newblock \eprint{1507.04627}.

\bibitem{Bazavov:2017dus}
A.~Bazavov \emph{et~al.},
\newblock \emph{{The QCD Equation of State to $\mathcal{O}(\mu_B^6)$ from
  Lattice QCD}},
\newblock Phys. Rev. D \textbf{95}(5), 054504 (2017),
\newblock \doi{10.1103/PhysRevD.95.054504},
\newblock \eprint{1701.04325}.

\bibitem{Mogliacci:2013mca}
S.~Mogliacci, J.~O. Andersen, M.~Strickland, N.~Su and A.~Vuorinen,
\newblock \emph{{Equation of State of hot and dense QCD: Resummed perturbation
  theory confronts lattice data}},
\newblock JHEP \textbf{12}, 055 (2013),
\newblock \doi{10.1007/JHEP12(2013)055},
\newblock \eprint{1307.8098}.

\bibitem{Haque:2014rua}
N.~Haque, A.~Bandyopadhyay, J.~O. Andersen, M.~G. Mustafa, M.~Strickland and
  N.~Su,
\newblock \emph{{Three-loop HTLpt thermodynamics at finite temperature and
  chemical potential}},
\newblock JHEP \textbf{05}, 027 (2014),
\newblock \doi{10.1007/JHEP05(2014)027},
\newblock \eprint{1402.6907}.

\bibitem{Giordano:2019slo}
M.~Giordano and A.~P\'asztor,
\newblock \emph{{Reliable estimation of the radius of convergence in finite
  density QCD}},
\newblock Phys. Rev. D \textbf{99}(11), 114510 (2019),
\newblock \doi{10.1103/PhysRevD.99.114510},
\newblock \eprint{1904.01974}.

\bibitem{Giordano:2019gev}
M.~Giordano, K.~Kap{\'a}s, S.~D. Katz, D.~N{\'o}gr{\'a}di and A.~P{\'a}sztor,
\newblock \emph{{Radius of convergence in lattice QCD at finite $\mu_B$ with
  rooted staggered fermions}},
\newblock Phys. Rev. D \textbf{101}(7), 074511 (2020),
\newblock \doi{10.1103/PhysRevD.101.074511},
\newblock \eprint{1911.00043}.

\bibitem{Mondal:2021jxk}
S.~Mondal, S.~Mukherjee and P.~Hegde,
\newblock \emph{{Lattice QCD Equation of State for Nonvanishing Chemical
  Potential by Resumming Taylor Expansion}} (2021), \eprint{2106.03165}.

\bibitem{Dimopoulos:2021vrk}
P.~Dimopoulos, L.~Dini, F.~Di~Renzo, J.~Goswami, G.~Nicotra, C.~Schmidt,
  S.~Singh, K.~Zambello and F.~Ziesch\'e,
\newblock \emph{{A contribution to understanding the phase structure of strong
  interaction matter: Lee-Yang edge singularities from lattice QCD}}  (2021),
\newblock \eprint{2110.15933}.

\bibitem{Lombardo:2005ks}
M.~P. Lombardo,
\newblock \emph{{Series representation: Pade' approximants and critical
  behavior in QCD at nonzero T and mu}},
\newblock PoS \textbf{LAT2005}, 168 (2006),
\newblock \doi{10.22323/1.020.0168},
\newblock \eprint{hep-lat/0509181}.

\bibitem{Pasztor:2020dur}
A.~P\'asztor, Z.~Sz\'ep and G.~Mark\'o,
\newblock \emph{{Apparent convergence of Pad\'e approximants for the crossover
  line in finite density QCD}},
\newblock Phys. Rev. D \textbf{103}(3), 034511 (2021),
\newblock \doi{10.1103/PhysRevD.103.034511},
\newblock \eprint{2010.00394}.

\bibitem{Borsanyi:2021sxv}
S.~Bors\'anyi, Z.~Fodor, J.~N. G{\"u}nther, R.~Kara, S.~D. Katz, P.~Parotto,
  A.~P\'asztor, C.~Ratti and K.~K. Szab\'o,
\newblock \emph{{Lattice QCD equation of state at finite chemical potential
  from an alternative expansion scheme}},
\newblock Phys. Rev. Lett. \textbf{126}(23), 232001 (2021),
\newblock \doi{10.1103/PhysRevLett.126.232001},
\newblock \eprint{2102.06660}.

\bibitem{Parisi:1983mgm}
G.~Parisi,
\newblock \emph{{ON COMPLEX PROBABILITIES}},
\newblock Phys. Lett. B \textbf{131}, 393 (1983),
\newblock \doi{10.1016/0370-2693(83)90525-7}.

\bibitem{Aarts:2009uq}
G.~Aarts, E.~Seiler and I.-O. Stamatescu,
\newblock \emph{{The Complex Langevin method: When can it be trusted?}},
\newblock Phys. Rev. D \textbf{81}, 054508 (2010),
\newblock \doi{10.1103/PhysRevD.81.054508},
\newblock \eprint{0912.3360}.

\bibitem{Sexty:2013ica}
D.~Sexty,
\newblock \emph{{Simulating full QCD at nonzero density using the complex
  Langevin equation}},
\newblock Phys. Lett. B \textbf{729}, 108 (2014),
\newblock \doi{10.1016/j.physletb.2014.01.019},
\newblock \eprint{1307.7748}.

\bibitem{Hasenfratz:1991ax}
A.~Hasenfratz and D.~Toussaint,
\newblock \emph{{Canonical ensembles and nonzero density quantum
  chromodynamics}},
\newblock Nucl. Phys. B \textbf{371}, 539 (1992),
\newblock \doi{10.1016/0550-3213(92)90247-9}.

\bibitem{Fodor:2001au}
Z.~Fodor and S.~D. Katz,
\newblock \emph{{A New method to study lattice QCD at finite temperature and
  chemical potential}},
\newblock Phys. Lett. B \textbf{534}, 87 (2002),
\newblock \doi{10.1016/S0370-2693(02)01583-6},
\newblock \eprint{hep-lat/0104001}.

\bibitem{Fodor:2004nz}
Z.~Fodor and S.~D. Katz,
\newblock \emph{{Critical point of QCD at finite T and mu, lattice results for
  physical quark masses}},
\newblock JHEP \textbf{04}, 050 (2004),
\newblock \doi{10.1088/1126-6708/2004/04/050},
\newblock \eprint{hep-lat/0402006}.

\bibitem{Giordano:2020uvk}
M.~Giordano, K.~Kap{\'a}s, S.~D. Katz, D.~N{\'o}gr{\'a}di and A.~P{\'a}sztor,
\newblock \emph{{Effect of stout smearing on the phase diagram from
  multiparameter reweighting in lattice QCD}},
\newblock Phys. Rev. D \textbf{102}(3), 034503 (2020),
\newblock \doi{10.1103/PhysRevD.102.034503},
\newblock \eprint{2003.04355}.

\bibitem{Fodor:2007vv}
Z.~Fodor, S.~D. Katz and C.~Schmidt,
\newblock \emph{{The Density of states method at non-zero chemical potential}},
\newblock JHEP \textbf{03}, 121 (2007),
\newblock \doi{10.1088/1126-6708/2007/03/121},
\newblock \eprint{hep-lat/0701022}.

\bibitem{Endrodi:2018zda}
G.~Endr{\H o}di, Z.~Fodor, S.~D. Katz, D.~Sexty, K.~K. Szab{\'o} and
  C.~T{\"o}r{\"o}k,
\newblock \emph{{Applying constrained simulations for low temperature lattice
  QCD at finite baryon chemical potential}},
\newblock Phys. Rev. D \textbf{98}(7), 074508 (2018),
\newblock \doi{10.1103/PhysRevD.98.074508},
\newblock \eprint{1807.08326}.

\bibitem{deForcrand:2002pa}
P.~de~Forcrand, S.~Kim and T.~Takaishi,
\newblock \emph{{QCD simulations at small chemical potential}},
\newblock Nucl. Phys. B Proc. Suppl. \textbf{119}, 541 (2003),
\newblock \doi{10.1016/S0920-5632(03)80451-6},
\newblock \eprint{hep-lat/0209126}.

\bibitem{Allton:2002zi}
C.~R. Allton, S.~Ejiri, S.~J. Hands, O.~Kaczmarek, F.~Karsch, E.~Laermann,
  C.~Schmidt and L.~Scorzato,
\newblock \emph{{The QCD thermal phase transition in the presence of a small
  chemical potential}},
\newblock Phys. Rev. D \textbf{66}, 074507 (2002),
\newblock \doi{10.1103/PhysRevD.66.074507},
\newblock \eprint{hep-lat/0204010}.

\bibitem{Morningstar:2003gk}
C.~Morningstar and M.~J. Peardon,
\newblock \emph{{Analytic smearing of SU(3) link variables in lattice QCD}},
\newblock Phys. Rev. D \textbf{69}, 054501 (2004),
\newblock \doi{10.1103/PhysRevD.69.054501},
\newblock \eprint{hep-lat/0311018}.

\bibitem{STAR:2017sal}
L.~Adamczyk \emph{et~al.},
\newblock \emph{{Bulk Properties of the Medium Produced in Relativistic
  Heavy-Ion Collisions from the Beam Energy Scan Program}},
\newblock Phys. Rev. C \textbf{96}(4), 044904 (2017),
\newblock \doi{10.1103/PhysRevC.96.044904},
\newblock \eprint{1701.07065}.

\bibitem{STAR:2020tga}
J.~Adam \emph{et~al.},
\newblock \emph{{Nonmonotonic Energy Dependence of Net-Proton Number
  Fluctuations}},
\newblock Phys. Rev. Lett. \textbf{126}(9), 092301 (2021),
\newblock \doi{10.1103/PhysRevLett.126.092301},
\newblock \eprint{2001.02852}.

\end{thebibliography}
